\documentclass[9pt,twocolumn,twoside]{osajnl}

\journal{ol} 

\setboolean{shortarticle}{false}

\ifthenelse{\boolean{shortarticle}}{\colorlet{color2}{color2b}}{\colorlet{color2}{color2}} 

\title{1.25 GHz sine wave gating InGaAs/InP single-photon detector with monolithically integrated readout circuit}

\author[1,2]{Wen-Hao Jiang}
\author[3]{Jian-Hong Liu}
\author[4]{Yin Liu}
\author[1,2]{Ge Jin}
\author[1,2,*]{Jun Zhang}
\author[1,2]{Jian-Wei Pan}

\affil[1]{Hefei National Laboratory for Physical Sciences at the Microscale and Department of Modern Physics, University of Science and Technology of China, Hefei, Anhui 230026, China}
\affil[2]{CAS Center for Excellence and Synergetic Innovation Center in Quantum Information and Quantum Physics, University of Science and Technology of China, Hefei, Anhui 230026, China}
\affil[3]{Quantum CTek Co., Ltd., Hefei, Anhui 230088, China}
\affil[4]{Yun Micro Electronics Co., Ltd., Hefei, Anhui 230094, China}
\affil[*]{Corresponding author: zhangjun@ustc.edu.cn}

\dates{Compiled \today}

\ociscodes{(270.5570) Quantum detectors; (040.1345) Avalanche photodiodes (APDs); (270.5568) Quantum cryptography.}

\doi{\url{http://dx.doi.org/10.1364/ao.XX.XXXXXX}}

\begin{abstract}
InGaAs/InP single-photon detectors (SPDs) are the key devices for applications requiring near-infrared single-photon detection.
Gating mode is an effective approach to synchronous single-photon detection. Increasing gating frequency and reducing module size are important challenges for the design of such detector system. Here we present for the first time an InGaAs/InP SPD with 1.25 GHz sine wave gating using a monolithically integrated readout circuit (MIRC). The MIRC has a size of 15 mm $\times$ 15 mm and implements the miniaturization of avalanche extraction for high-frequency sine wave gating. In the MIRC, low-pass filters and a low-noise radio frequency amplifier are integrated based on the technique of low temperature co-fired ceramic, which can effectively reduce the parasitic capacitance and extract weak avalanche signals. We then characterize the InGaAs/InP SPD to verify the functionality and reliability of MIRC, and the SPD exhibits excellent performance with 27.5 \% photon detection efficiency, 1.2 kcps dark count rate, and 9.1 \% afterpulse probability at 223 K and 100 ns hold-off time. With this MIRC, one can further design miniaturized high-frequency SPD modules that are highly required for practical applications.
\end{abstract}

\setboolean{displaycopyright}{true}

\begin{document}

\maketitle
\thispagestyle{fancy}

\ifthenelse{\boolean{shortarticle}}{\ifthenelse{\boolean{singlecolumn}}{\abscontentformatted}{\abscontent}}{}

InGaAs/InP single-photon detector (SPD), composed of a single-photon avalanche diode (SPAD) and quenching electronics, is widely used in practical applications requiring near-infrared single-photon detection such as quantum key distribution, lidar, luminescence detection and optical time domain reflectometer~\cite{ZIZ15}.
InGaAs/InP SPAD can be operated either in free-running mode or in gating mode, aiming for asynchronous and synchronous single-photon detections, respectively.

Free-running mode is suited for the applications in which photon arrival times are unknown. So far, various approaches have been implemented for free-running single-photon detection including passive quenching~\cite{RWR00,WIB09}, passive quenching and active reset (PQAR) scheme~\cite{PQAR08,PQAR10}, integrated circuit of active quenching~\cite{ASIC07,ASIC09},
negative feedback avalanche diodes~\cite{IJN09,YHH12,KWL14,ABK16,YC17}. However, in the free-running schemes relatively long hold-off time settings are often required to further reduce the afterpulse probability, which may limit the maximum count rate of InGaAs/InP SPDs.

For the applications in which photon arrival times are known, using gating mode can suppress both dark count rate (DCR) and afterpulse probability ($P_{ap}$).
The challenge in this scheme is to extract avalanche signals superimposed on derivative signals due to the capacitive responses of SPAD. Gating frequency is one of the most important parameters, which determines the maximum count rate of SPD. In the early stage, gating frequencies were limited at the level of 10 MHz. After the invention of high-frequency gating techniques including sine wave gating (SWG)~\cite{NSI06} and self-differencing~\cite{SD07}, gating frequency has been rapidly increased up to GHz. In high-frequency gating schemes,
ultrashort gating time extremely limits the quantity of charge created during avalanche process, which, therefore, results in effective suppression of the afterpulsing effect
and significant increase of count rate. As a consequence of high-frequency gating, the amplitudes of avalanche signals
are pretty small, i.e., at the level of mV. Therefore, the primary goal of readout circuit in the high-frequency gating schemes is to extract weak avalanche signals from large capacitive responses.

\begin{figure*}[htbp]
\centering
\includegraphics[width=18 cm]{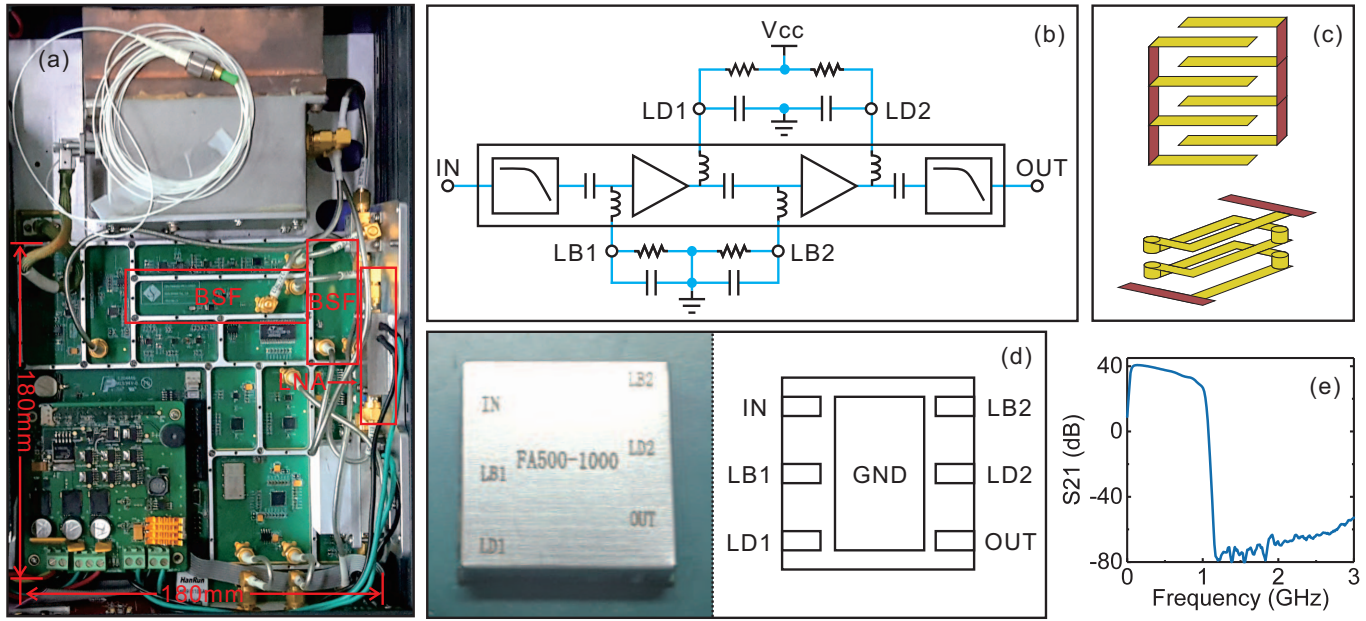}\\
\caption{(a) Photo of the conventional 1.25 GHz SWG SPD module with board-level integration designed in 2012. The size of avalanche extraction circuit marked by red rectangular regions is $\sim$ 5000 $mm^{2}$. (b) The functional block diagram and power configuration of MIRC. (c) The structures of LTCC capacitors (top) and inductors (bottom). (d) The photo (left) and pin configuration (right) of MIRC. (e) The measured S21 parameter of MIRC.}
\label{fig1}
\end{figure*}

For instance, in SWG schemes, sine waves with large amplitude are used as gates. Due to the frequency spectrum characteristic of sine waves, the capacitive responses of SPAD only include the fundamental frequency and higher order harmonics of sine waves. These response signals can be eliminated by filters. After the process of filtering-amplification-filtering in the readout circuit, weak avalanches can be finally extracted. So far many SWG schemes have been presented~\cite{NSI06,NAI09,ZTB09,GAP10,RGL11,NEC11,LLW12,WLG12,NIST13}.

The standard SWG readout circuit includes band-stop filters (BSFs) and a low-noise amplifier (LNA). Integration is crucial for detector system design, which can reduce both the size of SPD module and the parasitic capacitance as well. The parasitic capacitance reduction can effectively decrease the charge carriers during avalanche process and thus the afterpulse probability. Previously, the readout circuit was often implemented using discrete electronic components. In 2012, the stand-alone SWG InGaAs/InP SPD module with board-level integration was first reported~\cite{LLW12}, as shown in Fig.~\ref{fig1}(a). The SPD module with 1.25 GHz gating frequency integrated diverse functionalities, including automatic gain control for gate amplitude, low-noise tunable bias, automatic delay adjustment, avalanche signal extraction and monitor, tunable SPAD temperature, programmable discrimination threshold and hold-off time, synchronized clock outputs, and user-friendly interface~\cite{LLW12}. The total size of avalanche extraction circuit was $\sim$ 5000 $mm^{2}$, as marked by three red rectangular regions in Fig.~\ref{fig1}(a).
In order to significantly reduce the size of SWG SPD modules that are highly required for practical applications, developing monolithically integrated readout circuit is necessary.

\begin{figure}[htbp]
\centering
\includegraphics[width=9 cm]{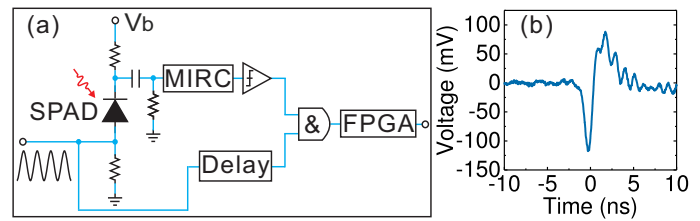}\\
\caption{(a) InGaAs/InP SPD using the MIRC. (b) Typical avalanche signal trace captured at the output of MIRC.}
\label{fig2}
\end{figure}

In this Letter, we present for the first time an InGaAs/InP single-photon detector using monolithically integrated readout circuit (MIRC) with 1.25 GHz clock rate. The functional block diagram of MIRC is depicted in Fig.~\ref{fig1}(b). The MIRC includes two low-pass filters (LPFs) and a two-stage low-noise radio frequency amplifier, which is biased via external resistors and capacitors. The LPFs are designed using miniaturized capacitors and inductors based on the technology of low temperature co-fired ceramics (LTCC), whose structures are shown in Fig.~\ref{fig1}(c). The LPF has a cut-off frequency of $\sim$1 GHz and a rejection ratio of $\sim$ 60 dB at 1.25 GHz.

LTCC technology can provide monolithic and ceramic microelectronic devices using multilayer printed circuit board (PCB) like structures, which, therefore, brings excellent radio frequency performance. The miniaturized LPFs are fabricated using the standard LTCC process. First, via holes are punched into the green ceramic tapes and filled with silver paste, which enables the electrical connections between layers. Meanwhile, the circuits are printed onto the tapes. Then, the tapes are stacked and laminated together using an isostatic press, and the laminated tapes are cut into desired shapes. Finally, the silver paste and green ceramic tapes are co-fired together at $\sim$ 900 $^\circ$C.

The MIRC is wholly packaged in a size of 15 mm $\times$ 15 mm, as shown in Fig.~\ref{fig1}(d), so that the size ratio of MIRC to the avalanche extraction circuit
inside the conventional SWG SPD module designed in 2012 is only 4.5\%.
Before applying this MIRC into a SWG SPD, we first characterize the S21 parameter of MIRC using a network analyzer. The result is shown in Fig.~\ref{fig1}(e), from which one can observe that the frequency components below $\sim$ 1 GHz are amplified with a gain of 40 dB roughly while the center frequency of 1.25 GHz is rejected with $\sim$ 80 dB. Therefore, using this MIRC weak avalanche signals can be easily extracted.

\begin{figure}[tbp]
\centering
\includegraphics[width=7.1 cm]{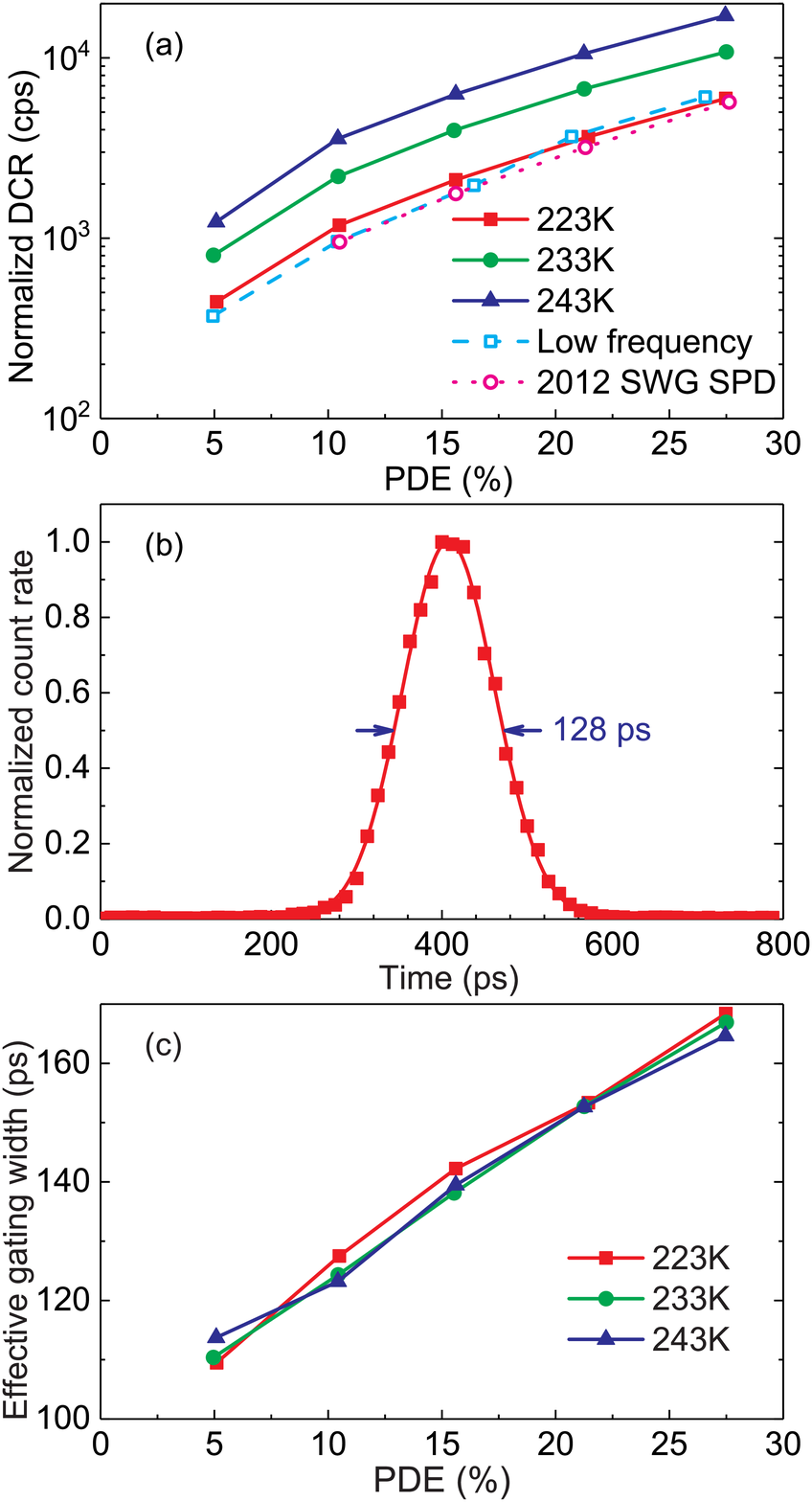}\\
\caption{(a) DCR normalized by duty cycle versus PDE. Solid lines show the measured results using MIRC with three temperature settings.
Dotted and dashed lines show the results at 223 K using the conventional SWG SPD and standard low-frequency characterization electronics, respectively. (b) Effective gating width measurement by scanning the delay between laser pulses and gates at 10\% PDE and 223 K. (c) Effective gating width versus PDE with three temperature settings.}
\label{fig3}
\end{figure}

We then develop a MIRC-SWG InGaAs/InP SPD with 1.25 GHz gating frequency, as shown in Fig.~\ref{fig2}(a). In the experiment, we use the same circuit as in our previous work~\cite{LLW12} for SPAD temperature control, sine wave generation, avalanche discrimination and automatic delay adjustment.
The SPAD cathode is biased with a voltage of $V_{b}$ through a resistor of 1 K$\Omega$. Sine waves with $\sim$11 V peak-peak amplitude are gated on the SPAD anode. The capacitive responses from the cathode are alternating current coupled to the input of MIRC, which directly extracts avalanche signals. Fig.~\ref{fig2}(b) shows a typical oscilloscope trace of avalanche signal, which exhibits high signal-to-noise ratio.

The avalanche signals are entered into a comparator, and coincidence is performed between the discriminated signals and synchronized signals of gates to remove the distorted output signals occurred out of gate duration. The relative delay is tuned by a precise phase shifter. A field-programmable gate array (FPGA) is used to set hold-off time, or count-off time, to further suppress the afterpulse probability. The approach of count-off time was proposed in 2009~\cite{ZTB09}. The MIRC-SWG SPD continuously runs, and given a triggered avalanche signal the avalanches during the following hold-off time are not recorded. Previous experiments have exhibited the usefulness of this approach to suppress $P_{ap}$ for SWG schemes~\cite{ZTB09,GAP10,LLW12}.

Further, we characterize the key parameters of MIRC-SWG SPD (SPAD model: PGA-300, Princeton Lightwave) to verify the functionality of MIRC.
Fig.~\ref{fig3}(a) plots the relationship between DCR that is normalized by duty cycle and photon detection efficiency (PDE) at 223 K, 233 K and 243 K, respectively. At 10\% PDE and 223 K, measured DCR is 188 cps or 1.5$\times10^{-7}$/gate. Considering the duty cycle due to the effective gating width ($\Delta t$) as shown in Fig.~\ref{fig3}(b), the normalized DCR is $\sim$ 1.18 kcps. As PDE increases to 27.5\% at 223 K, measured DCR and normalized DCR exponentially increase to 1.2 kcps and 5.96 kcps, respectively.

For comparison, we measure the relationship between DCR and PDE for the same SPAD at 223 K using both the conventional SWG SPD module~\cite{LLW12} and a standard low-frequency characterization electronics~\cite{ZIZ15} (10 kHz gating frequency and 50 ns gating width), as shown in the dotted and dashed lines of Fig.~\ref{fig3}(a), respectively. The three lines almost overlap each other, which clearly indicates that DCR, as an intrinsic parameter of SPAD, is independent of quenching electronics. Fig.~\ref{fig3}(c) shows that $\Delta t$ is roughly a linear function of PDE and this linear trend is independent of temperature.

\begin{figure}[htbp]
\centering
\includegraphics[width=7.5 cm]{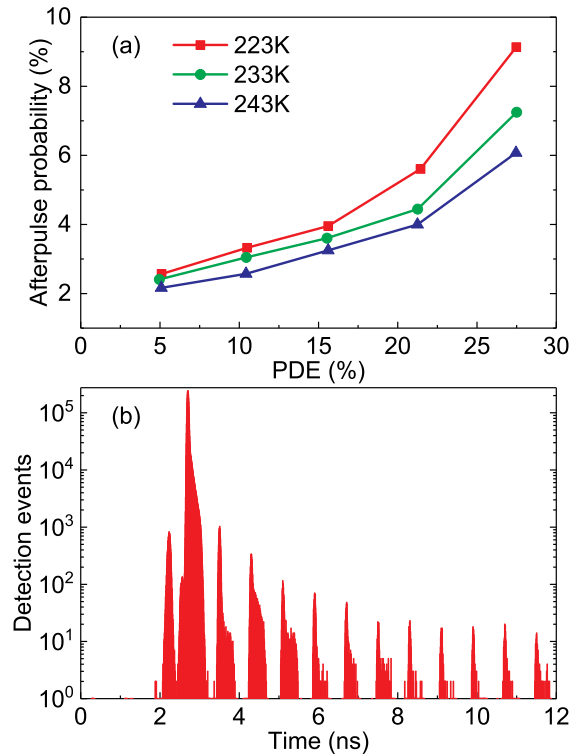}\\
\caption{(a) $P_{ap}$ versus PDE.(b) The histogram of detection events under the conditions of 10\% PDE, 223 K and 100 ns hold-off time.}
\label{fig4}
\end{figure}

Afterpulse probability is another important parameter of InGaAs/InP SPD.
Fig.~\ref{fig4}(a) plots the measured results of $P_{ap}$ versus PDE with three temperature settings.
We use the same method as in the reference~\cite{LLW12} to calculate $P_{ap}$.
We first measure the detection event distribution using a time-to-digital converter (PicoHarp 300, PicoQuant).
Fig.~\ref{fig4}(b) shows a typical histogram of detection distribution under the conditions of 10\% PDE, 223 K, 100 ns hold-off time, laser frequency of 625 kHz and mean photon number per pulse of 1. After subtracting the approximately uniform DCR contribution, the afterpulsing distribution is obtained, from which photon detection counts and afterpulse counts are summed, respectively~\cite{LLW12}.
As temperature increases, $P_{ap}$ rapidly decreases, particularly in the case of high PDE. For instance, at 27.5\% PDE the values of $P_{ap}$ are 9.1\% , 7.2\%, and 6.1\% at 223 K, 233 K, and 243 K, respectively.

One can compare DCR and afterpulse probability in terms of noise contribution. For instance, under the conditions of 10\% PDE, 223 K and 100 ns hold-off time, $P_{ap}$ is measured as 3.3\% over 16 $\mu$s, corresponding to 1.6$\times10^{-6}$/gate, which is higher than the value of DCR with one order of magnitude. Using long hold-off time, $P_{ap}$ can be significantly reduced down to the same level as DCR.

\begin{figure}[tbp]
\centering
\includegraphics[width=7.5 cm]{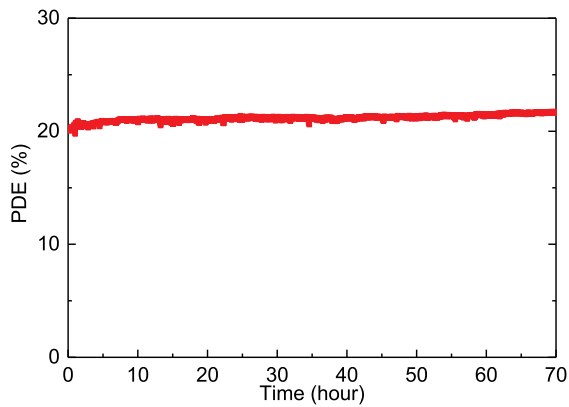}\\
\caption{Stability test result of MIRC-SWG SPD. The data during the delay scanning process are not plotted.}
\label{fig5}
\end{figure}

Finally, we test the stability of MIRC-SWG SPD. The PDE is initially set to $\sim$ 20\% at 223 K. During the test, every 10 minutes the delay between laser pulses and sine wave gates is scanned in the full range and set to the point of peak count, and every 1 minute detection counts with 10 s integration time are recorded. The stability test is continuously performed for over 70 hours. The test results are shown in Fig.~\ref{fig5}, from which one can conclude that the MIRC-SWG SPD is highly stable and reliable.

This MIRC can be further used to design a miniaturized all-in-one SWG SPD module. According to our estimation, the SPAD device, MIRC and mini thermoelectric cooler can be encapsulated inside a butterfly package with a size of 42 mm $\times$ 21 mm $\times$ 9 mm. Considering the necessary affiliated circuits including sine wave gate generation, temperature control, avalanche discrimination, automatic delay adjustment, FPGA and power supply, we evaluate that the miniaturized all-in-one SWG SPD module can be designed within a size of $\sim$12 cm $\times$ 7 cm $\times$ 5 cm. Compared with the conventional SWG SPD module ($\sim$25 cm $\times$ 10 cm $\times$33 cm)~\cite{LLW12}, the module size is extremely reduced by 95\%, which is favorable for practical applications.

In conclusion, we report for the first time a 1.25 GHz sine wave gating InGaAs/InP SPD using monolithically integrated readout circuit. The chip with 15 mm $\times$ 15 mm in size, fabricated based on the LTCC technology, can directly extract avalanche signals. Moreover, such integration of readout circuit effectively reduces both the size and the parasitic capacitance. We further experimentally characterize the SPD and perform stability test to verify the functionality and reliability of MIRC. Using such MIRC, one may design a miniaturized SWG SPD module with considerably reduced size.

\section*{Funding Information}

National Key R\&D Program of China (2017YFA0304004); National Natural Science Foundation of China (11674307); Chinese Academy of Sciences.






\end{document}